# Characterization and diagnostics of combustion thermoacoustic instabilities using nonlinear dynamics and topological methods

G. Gianni, G. Mariotti, E. Paganini, S. Sello[*]
Mathematical and Physical Models
Enel Research
Pisa, IT

**Abstract**
The onset of thermoacoustic instabilities in lean-premixed gas-turbine combustors is a crucial problem leading to degradation in engine and emissions performance and shortened component life. The main aim of this study is to propose a methodology based both on concepts of nonlinear dynamics and on geometric-topological invariants, for the characterization of attractors related to measurements based on the flame spontaneous light emission, like OH* radical, in order to classify different phases of the combustion process and to better recognize the transition mechanisms leading to the thermoacoustic instabilities. Preliminary results, clearly show the powerfulness of the approach to show the dynamical evolution of the flame and to evidence the onset of the thermoacoustic instabilities: in particular the topological invariant index (genus and related quantities) appears as the best candidate for an early indicator of the dynamical transition, characterized by the onset of a more persistent, low entropy torus (quasi-periodic motion), as the related "thermoacoustic attractor".

**Introduction**

The investigation of combustion instability mechanisms in lean premixed gas turbine combustors is of great interest due to increasingly stringent emissions demands, particularly in regards to NOx emissions. These systems produce significantly lower amounts of emissions than conventional gas turbine combustors by burning the fuel at low temperatures. On the other hand, these operating conditions are characterized by the occurrence of combustion instabilities, which are mainly manifested as large amplitude, organized oscillations of the flow field excited and coupled by unsteady combustion processes. Such combustion instabilities can cause a significant degradation of the combustion performances and a reduction of components life.

The main aim of this study is to propose a methodology, based both on concepts of nonlinear dynamics and on geometric-topological invariants, for the characterization of attractors related to high frequency measurements of the flame spontaneous light emission, like OH* radical. This approach allows an accurate classification of different phases of the combustion process and a better recognition of the *transition mechanisms* leading to the thermoacoustic instabilities. The nature of the mechanism for the onset of the self-sustained interaction between pressure and flame for different operating conditions is not yet clear. Some authors assume that this process is always stochastical in the beginning.[1] Another possibility is that the process can be described by a complex nonlinear deterministic system with the ability of an early diagnostics. The method here proposed can help to shed more light on this topic.

**Specific Objectives**

The main objective of the present investigation is a deep phenomenological characterization of the combustion thermoacoustic instability as a dynamical state, including the related onset phase, in a scaled lean premixed combustor utilizing analyses of pressure fluctuations and optical (radical OH*) flame emissions time series, based both on wavelet techniques and on nonlinear dynamics tools, like suitable geometric and topological invariants.

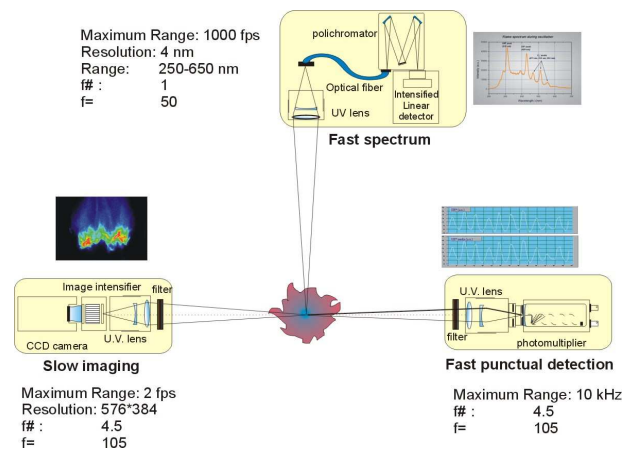

Fig. 1. Experimental setup: measurement instruments.

Experimental Setup

The experimental technique used to record chemiluminescence measurements from the flame is based on a filtered high frequency and high sensitivity photomultiplier giving a 10 kHz sampling frequency in order to reach the required time resolution. The experimental set-up is installed on a scaled combustion chamber at atmospheric pressure, equipped with quartz windows to have direct optical access to the flame. The spatial structure and time evolution of the flame is recorded through a high frequency and intensified camera. The combustion system can be operated at the maximum power of 400 kW. The fuel is natural gas, fed through three different lines (pilot, diffusion, premix). See Fig.1.[2]

Methods of Analysis

The first method used to characterize the unsteady OH* and pressure time series is the *wavelet analysis*. Fourier analysis is in fact an adequate tool for detecting and quantifying constant periodic fluctuations in time series. For intermittent and transient multiscale phenomena, the wavelet transform is able to detect time evolutions of the frequency distribution. The continuous wavelet transform represents an optimal localized decomposition of time series, $x(t)$, as a function of both time $t$ and frequency (scale) $a$, from a convolution integral:[3]

$$W(a,\tau) = \frac{1}{a^{1/2}} \int_{-\infty}^{+\infty} dt\, x(t) \psi^*\left(\frac{t-\tau}{a}\right)$$

where $\psi$ is called an analysing wavelet if it verifies the following admissibility condition:

$$c_\psi = \int_0^{+\infty} d\omega\, \omega^{-1} \left|\hat{\psi}(\omega)\right|^2 < \infty$$

where:

$$\hat{\psi}(\omega) = \int_{-\infty}^{+\infty} dt\, \psi(t) e^{-i\omega t}$$

is the related Fourier transform. In the definition, $a$ and $\tau$ denote the dilation (scale factor) and translation (time shift parameter), respectively.

We define the local wavelet spectrum:

$$P_\omega(k,t) = \frac{1}{2 c_\psi k_0} \left|W\left(\frac{k_0}{k}, t\right)\right|^2 \quad k \geq 0$$

where $k_0$ denotes the peak frequency of the analysing wavelet $\psi$.

From the local wavelet spectrum we can derive a mean or global wavelet spectrum, $P_\omega(k)$:

$$P_\omega(k) = \int_{-\infty}^{+\infty} dt\, P_\omega(k,t)$$

The relationship between the ordinary Fourier spectrum $P_F(\omega)$ and the mean wavelet spectrum $P_\omega(k)$ is given by:

$$P_\omega(k) = \frac{1}{c_\psi k} \int_0^{+\infty} d\omega\, P_F(\omega) \left|\hat{\psi}\left(\frac{k_0 \omega}{k}\right)\right|^2$$

indicating that the mean wavelet spectrum is the average of the Fourier spectrum weighted by the square of the Fourier transform of the analysing wavelet $\psi$ shifted at frequency $k$.

Here we used the family of complex analysing wavelets consisting of a plane wave modulated by a Gaussian (called Morlet wavelet):[4]

$$\psi(\eta) = \pi^{-1/4} e^{i\omega_0 \eta} e^{-\eta^2/2}$$

where $\omega_0$ is the non dimensional frequency here taken to be equal to 6 in order to satisfy the admissibility condition.

Let $W_f(a,\tau)$ and $W_g(a,\tau)$ be respectively the wavelet transforms of two real signals: $f(t)$ and $g(t)$. We define the wavelet cross-scalogram as:[5]

$$W_{fg}(a,\tau) = W_f^*(a,\tau) W_g(a,\tau)$$

If the analyzing wavelet is complex the cross-scalogram is also complex and can be written in terms of its real and imaginary parts:

$$W_{fg}(a,\tau) = CoW_{fg}(a,\tau) + iQuadW_{fg}(a,\tau)$$

and it can be shown that the following relation holds:

$$\int_{-\infty}^{+\infty} dt\, f(t) g(t) = (1/c_\psi) \int_0^{\infty} d\tau \int_{-\infty}^{+\infty} da\, CoW_{fg}(a,\tau)$$

The wavelet cross-correlation analysis allows to extend the standard Fourier cross-correlation analysis to non-stationary signals whose power spectral characteristics are not constant in time.

This approach gives interesting results for the recognition of the states of dynamical systems but it cannot be used for an early diagnostic because state variations appear suddenly in the graphic. Moreover the long computational time required makes it inadequate to be used in real time. This brought us to study and to use methods based on nonlinear dynamic tools.

The first step is to consider the reconstruction of the attractor associated to the dynamical process in the Embedding Space, using the Embedding Theorem [6]. Then we make a geometrical-topological characterization of attractors associated to the phases of dynamical evolution using some dynamical indicators:

- Attractor entropy – a dispersion measure;
- Trajectory curvature – a geometrical index;
- Genus – a topological invariant index.

Given an attractor in $R^n$ associated to a dynamical system we define the attractor *entropy* as the Shannon Entropy:[7]

$$H : P \to [0, \infty)$$

$$H(X) = -\sum_{x \in dom(X)} f(x) \log_2 f(x)$$

where $X$ is an random variable, $P$ is the space of the distributions with finite probability:

$$P = \{f \mid f : dom(X) \to [0,1], x \in dom(X) = \{s_1, \ldots, s_n\}\}$$

where $n=1,2,...$, $dom(X)$ is the family of $X$'s values, $x$ is an $X$ value and $f$ is the probability distribution of $X$.

Entropy measures the dispersion of an attractor and the regularity of a process: a very regular process has a small attractor and low entropy, while an irregular one has a large attractor and high entropy.

The *curvature* for a curve $C$ parametrized in $t$ in a $n$-dimensional space:

$$\{x(t)\}=\{q(t),p(t)\}$$

is defined as:

$$c(t) = \left|\frac{dt}{ds}\right| = \left|\frac{d^2x}{ds^2}\right| \quad (1)$$

where $t=dx/ds$ is the unity tangent vector in the point and $s=s(t)$ is curvilinear coordinate:

$$s(t) = \int_0^t \sqrt{\sum_{k \in I}(dx_k/d\tau)^2}\, d\tau$$

We can write the curvature also as:

$$c(t) = \sqrt{\sum_{k \in I}(d^2x_k/ds^2)^2} \quad (2)$$

So $c$ is the angular deviation velocity of the unity tangent vector if the velocity on $C$ is unitary.
We calculate the curvature of an attractor using the discrete versions of (1) and (2) which has proven to be a good dynamic index [7].

The *genus* is a topological invariant which for a surface in a three dimensional space is defined as the number of holes minus the number of connected components of the surface.

Using the Gauss-Bonnet theorem we can calculate the genus using the gaussian curvature $K$:

$$G = -\frac{1}{4\pi}\int k\, dA$$

The use of a topological property to characterize the states of a dynamical process is justified by the observation of equivalence between dynamical and topological equivalence [9][10].

To calculate the genus of an attractor we use a discrete formula, valid for all the regular polygonal surfaces in a three dimensional space:

$$G = -\frac{1}{4\pi}\sum D_i \quad (3)$$
$$D_i = 2\pi - \sum V_i$$

where $V_i$ are the angles incident on the i-th vertex.

In fact, we implemented an algorithm which approximates the attractor surface with a polygonal one made by cubes and calculates the genus using the relation (3).

Our algorithm is based on that used by Gott et al. to determine the topological structure of the universe in [11] and described in [12].

We modified the original algorithm so that the fundamental parameters of the process, which are the length of the edge of every cube $\lambda$ and the critical density $\rho_c$, are chosen univocally using the characteristics of the attractor. So we choose $\lambda$ as the average distance between the points of the attractor $d_m$ plus its standard deviation $\sigma_d$:

$$\lambda = d_m + \sigma_d$$
$$\sigma_d = \sqrt{\frac{\sum(d_m - d_k)^2}{N-1}}$$

and $\rho_c$ as the average density of the cubes $\rho_m$ plus its standard deviation $\sigma_\rho$:

$$\rho_c = \rho_m + \sigma_\rho$$
$$\sigma_\rho = \sqrt{\frac{\sum(\rho_k - \rho_m)^2}{N-1}}$$

It has been experimentally verified that the thermoacoustic instability state is characterized by having toroidal attractors and hence genus 0. This isn't an unexpected result because this instability state can be seen as a quasi-periodic motion.

So from the genus of the attractors of a dynamical process we extract two further indices, useful to characterize the states of a process and to determine if they are stable or unstable.

The *persistency index* counts the percentage of attractors with genus 0 during the system evolution so tells us how near we are to the instability.

The *torus quality index, TQI,* measures the inverse of entropy every time the genus has 0 value. So it measures the "quality" (the dispersion properties) of the toroidal attractors and allow us to distinguish between phases with attractors with the same shape but different regularity.

In order to follow the temporal evolution of the above indices we used a moving time-window version of their mathematical formulation. In this way we were able to accurately detect the transition between different dynamical states.

**Results and Discussion**

The above methods were first tested on simple analytical signals to set up the available free parameters. In Fig. 2 we show an example of wavelet analysis performed on a real case in premixed conditions (see case A in Tab. 1), using an optical flame emission from radical OH*.

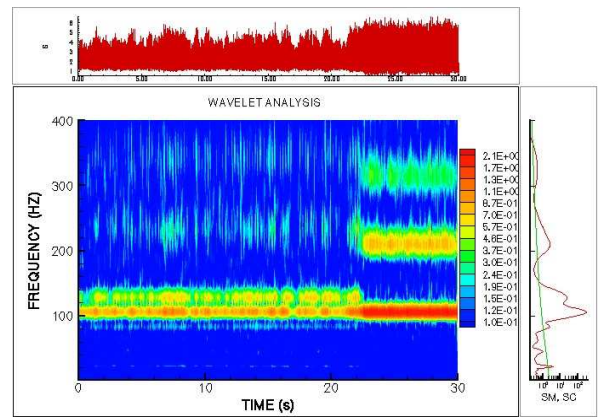

Fig. 2. Wavelet map for OH* signal (case A). Upper panel: original time series; central panel: wavelet contour map; right panel: global wavelet power (red line) and 95% confidence level for a background red noise process (green line).

**TABLE 1**
Summary of the principal process parameters for two different combustion cases.

| Case | Air flow rate g/s | T air C° | Pilot g/s | Fuel flow rate Diffusion g/s | Premix g/s | Φ | Pilot/total |
|---|---|---|---|---|---|---|---|
| A | 238→236 | 43 | 0.76 | 2.36→1.59 | 4.73→5.52 | 0.57→0.57 | 39.8→29.9 |
| B | 320→319 | 43 | 0.63→0.29 | 0 | 9.68→9.96 | 0.54→0.53 | 6.1→2.8 |

As indicated by process parameters displayed in Tab. 1, this time series is not stationary, with a time changing for both frequency and amplitude content. Through the wavelet analysis we are able to follow the evolution of the power spectrum: we start with a dominant characteristic frequency at 105 Hz and a coupled secondary frequency at 125 Hz. The wavelet analysis allows to point out the amplitude intermittency of the above frequencies, probably related to vortex shedding phenomena. At about 22 seconds there is a clear spontaneous transition to a thermoacoustic instability, evidenced by a strong rise of the emission amplitude with the onset of the next few harmonics, characteristic of the acoustic field of the combustion system considered. This fact is confirmed by the wavelet map of the pressure fluctuation signal shown in Fig. 3. Indeed, the signal amplitude and the wavelet power reach values typical for thermoacoustic oscillations.

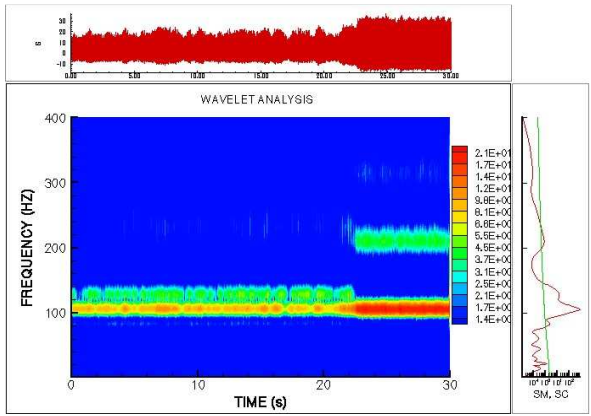

Fig. 4. Wavelet map for pressure fluctuation signal (case A).

The results of our non linear analysis are summarized by Figs. 4 and 5 for OH* emission and pressure respectively.

Before to analyze the signal we verified that the embedding dimension is equal to three, so that we could use the topological methods described before.

Observing Fig. 4 we can see that all the indices represented allow us to distinguish the different dynamical states and in particular the transition phase to the thermoacoustic instability, specially the entropy and the topological ones.

In fact, the entropy value makes oscillation near the value 7 during the phase of stability and near the value 5.5 during the instability one, with a finite time gradual transition between them. This is a first indication supporting the deterministic nature of the processes driving the transition to thermoacoustic oscillation state.

The curvature initially oscillates near the value 1.5, has some bigger oscillations during the transition phase and assumes in a almost stable manner the value 1.7 during the oscillating phase.

The genus initially oscillates near the 0 value and in the second phase is always equal to 0, even if the transition for this index seems to be more sudden and unpredictable.

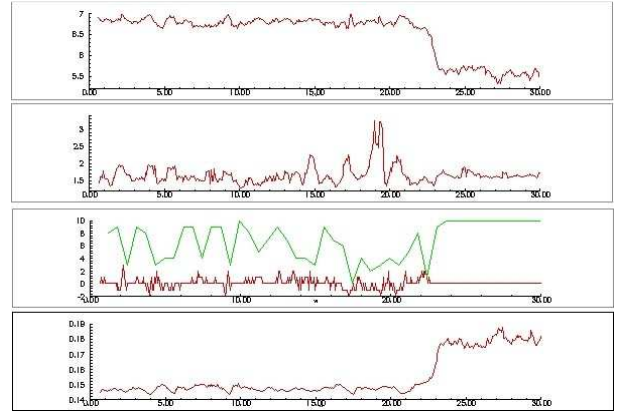

Fig. 3. Non linear analysis for OH* signal: geometrical and topological indices (case A). From top to bottom: Entropy, Mean Curvature, Topological Invariants (Genus: red line, Persistency Index: green line), TQI.

The persistency index instead oscillates and sometimes reaches its maximal value during the stable phase and is always equal to 10 during the unstable one, so with its behavior advises us about the possibility of thermoacoustic instability throughout the test. The behavior of this two indices confirms the fact that during the thermoacoustic instability the related attractor has always toroidal shape. This fact is consistent with a quasi-periodic motion characterizing the thermoacoustic oscillation (Fig. 5).

The torus quality index oscillates around the 0.15 value during the first phase and around 0.18 during the second, with a finite time gradual transition between them. From this we can deduce that although the attractors have often toroidal-like shape also during the non-oscillatory phase, we can distinguish between them considering their regularity: the torus of the oscillatory phase is more definite and less diffusive than that of the non-oscillatory state.
The nonlinear attractor characterization of an unsteady premixed flame tip motion of a rotating Bunsen flame, showed the possibility of a transition from a cycle limit, a torus like quasi-periodic motion and a non-periodic (chaotic) motion, depending on the flow velocity [13].

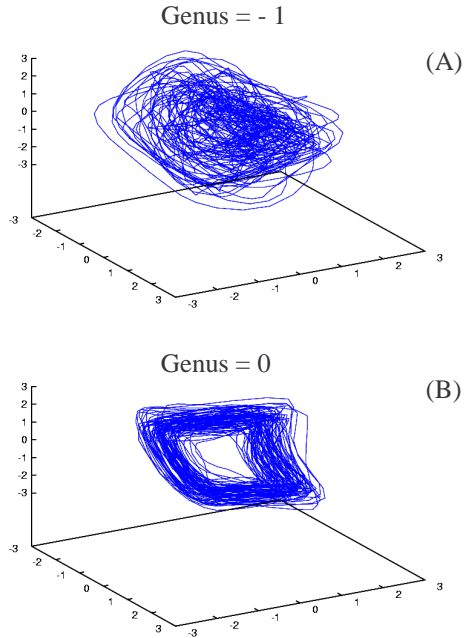

Fig. 5. Attractors in the embedding space: dynamical state distant from oscillating phase: spheroidal shape (A) and thermoacoustic phase: toroidal shape (B).

Similar results are obtained for pressure signal in Fig. 6, with the difference that in this case the genus and the quality index are less useful to recognize the transition. In fact we found transitions similar to those observed for OH* signal for the entropy, the curvature and the TQI, while the genus in this case is almost always equal to 0, so the persistency index is almost always equal to 10 and the difference between the two phases is more difficult to distinguish using them.

The optical signal appears a good choice to obtain accurate information about non stationary unstable combustion processes. However, the combination of optical and pressure signals is very interesting to know some important information about the amplitude-phase correlations. In Fig. 7 we show the wavelet cross-correlation analysis for the case A. The edge of the maximum correlation, indicated by the white line in the map, becomes very stable during the thermoacoustic phase. The related phase lag between the dominant mode of OH* and pressure signals changes toward higher absolute values during the transition and the average value in the thermoacoustic regime shows that the pressure signal precedes the optical one by 2.4 ms, which corresponds to an angular phase of 144°. This delay information can be very useful to set up accurate prediction models for combustion instabilities.

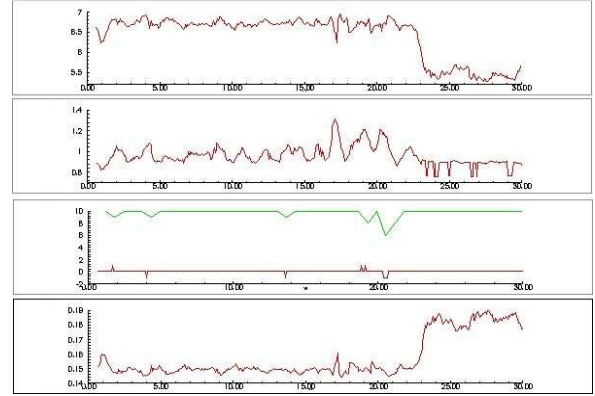

Fig. 6. Non linear analysis for pressure signal: geometrical and topological indices (case A). ). From top to bottom: Entropy, Mean Curvature, Topological Invariants (Genus: red line, Persistency Index: green line), TQI.

It is significant to observe that all the considered nonlinear indices allow us to recognize the transition between the two regimes, although some are more interesting if calculated using the OH* rather than pressure signal, and that all the transitions are synchronous. Moreover, we can observe that the transition happens in a finite time interval, about 2-3 seconds, thus the process that brings to instability seems to be predictable in advance, following a gradual deterministic behavior.

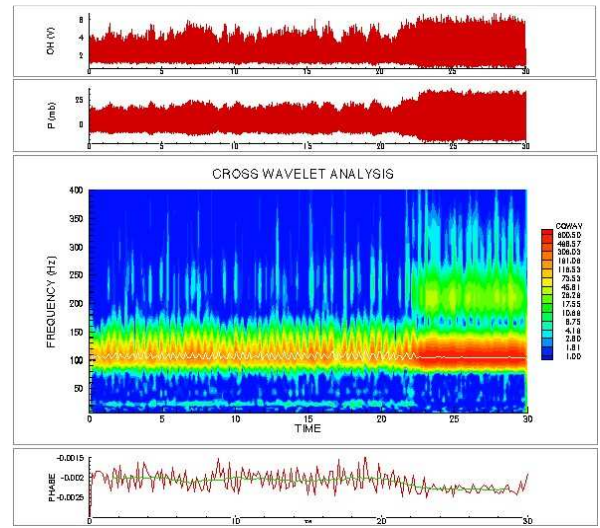

Fig. 7. Wavelet cross-correlation map for OH* and pressure signal (case A). White line is the edge maximum cross-correlation. The green line is the related phase lag function smoothed by a linear low-pass moving window filter.

The reason for preferring these nonlinear methods to more classical ones, including the wavelet analysis, is mainly in the fact that the non linear analysis allows to detect the transition in a more accurate level, requiring also a smaller computational time, so that they appear more useful for real time applications. Another important feature is that through the nonlinear analysis we can try to recognize the transition a little time before its complete development, so that we can use them to make, or design, an early diagnostic of the process. It is important to note that the entropy and curvature parameters can be applied also when the embedding space dimension is greater than three, probably a common situation in full scale combustion systems.

We selected a second combustion case, case B in Tab. 1, to check the performances of the nonlinear methods in a different flow configuration. In this case we eliminated the contribution of the main diffusion fuel flow rate.

The time behavior of the nonlinear indices presents here bigger fluctuations, because of the different conditions of the combustion process which make the interactions between flow and flame more unstable. However, the main findings of the previous case are confirmed. In fact, we find that during the instability state all the attractors have genus 0, confirming that they have low entropy toroidal shape with a quasi-periodic motion, and the transition occurs in a finite time interval, supporting the underlying deterministic evolution and feeding the hopes about its predictability.

**Conclusions**

The proposed methodology, based both on concepts of nonlinear dynamics and on geometric-topological invariants and supported by a wavelet analysis, allowed the characterization of the attractors related to measurements based on the flame spontaneous light emission, like OH* radical, and pressure fluctuations in a scaled lean premixed combustor. In order to classify different phases of the combustion process and to better recognize the transition mechanisms leading to the thermoacoustic instabilities, we used a moving time-window version of the dynamical indicators. Preliminary results clearly show that all the considered nonlinear parameters allow us to recognize the transition between the stable and unstable regimes, in particular when calculated using the OH* signal, and that all the transitions are synchronous. Moreover, we found that the transition happens in a finite time interval, about 2-3 seconds, i.e. the process that drives the instability seems to be predictable in advance, following a gradual deterministic behavior.

Moreover, the topological invariant index (genus and related quantities) clearly showed that the typical attractor of the thermoacoustic oscillation state is a persistent, low entropy torus describing a quasi-periodic motion.

We need further work, both analytical and experimental, to verify the reliability and the accuracy of the proposed methodology to full scale combustion systems with different process parameters.